\documentclass[11pt]{article} 
\usepackage{hyperref} \pdfoutput=1

\begin{document} 
\title{Leidenfrost explosions}
\author{Florian Moreau$^1$,  St\'ephane Dorbolo$^1$, Pierre Colinet$^2$ \\ 
\\\vspace{6pt}  $^1$GRASP, Physics Department, University of Liege, Belgium 
\\ $^2$Transfers, Interfaces and Processes (TIPs), Fluid Physics Unit 
\\Universit\'e Libre de Bruxelles, Belgium}

\maketitle
\begin{abstract} 
We present a fluid dynamics video showing the behavior of Leidenfrost
droplets composed by a mixture of water and surfactant (SDS,
Sodium Dodecyl sulfate).

When a droplet is released on a plate heated above a given temperature
a thin layer of vapor isolates the droplet from the plate. The
droplet levitates over the plate. This is called the Leidenfrost effect.

In this work we study the influence of the addition of a surfactant on
the Leidenfrost phenomenon. As the droplet evaporates the concentration
of SDS rises up to two orders of magnitude over the Critical Micelle
Concentration (CMC). An unexpected and violent explosive behavior is observed.
The video presents several explosions taken with a high speed camera
(IDT-N4 at 30000 fps). All the presented experiments were performed
on a plate heated at 300¡C. On the other hand, the initial
quantity of SDS was tuned in two ways: (i) by varying the initial concentration
of SDS and (ii) by varying the initial size of the droplet.
By measuring the volume of the droplet just before the explosion, we
were able to estimate the final concentration of SDS. We found that
the explosion always occurs around a critical concentration, about 100 times
the CMC.

The droplets have also been studied just before the explosion. By isolating
the droplet on a cold plate just before the explosion, we evidenced the
presence of a shell surrounding a liquid core.

We conclude that above a critical concentration a solid shell is
formed. This leads to an increase of pressure into the droplet until
the shell breaks. The release of the pressure is accompanied by a violent
explosion, and in some cases foaming.

\end{abstract}
\end{document}